\documentclass[a4paper, 12pt, openany, oneside]{article}
\usepackage[cp1251]{inputenc}
\usepackage[english]{babel}
\usepackage{ graphics,amsfonts, amsmath,bm,amssymb, array,}
\usepackage[dvips]{graphicx}
\makeatletter
\renewcommand{\Re}{\mathop{\rm Re\,}}

\makeatother {

\begin{document}

\thispagestyle{empty}
\renewcommand{\refname}{\begin{center} REFERENCES\end{center}}

 \begin{center}
\bf Interaction of Electromagnetic P--Wave with Metal
Films Located Between Two Dielectric Mediums
\end{center}\medskip
\begin{center}
  \bf  A. V. Latyshev\footnote{$avlatyshev@mail.ru$} and
  A. A. Yushkanov\footnote{$yushkanov@inbox.ru$}
\end{center}\medskip

\begin{center}
{\it Faculty of Physics and Mathematics,\\ Moscow State Regional
University, 105005,\\ Moscow, Radio str., 10--A}
\end{center}\medskip

\begin{abstract}
Generalization of the theory of interaction of the electromagnetic
$P$ -- waves  with a metal film on a case of the
film concluded between two dielectric environments is carried out.

\medskip

structures
\end{abstract}

\begin{center}
{\bf 1. Introduction}
\end{center}

The problem of interaction of an electromagnetic wave with a metal
film long time draws to itself attention \cite {F69} --
\cite {ly1}. It is connected as with theoretical interest to this
problem, and with numerous practical applications.

Nowadays there is the theory of interaction of an electromagnetic wave
 with metal film
in the case when electron reflection from a film surface
has a specular character \cite {F69} --
\cite {F09}.  In these works it was  considered the freely
hanging films in the air. In other words  it was considered the case, when
dielectric permeability of the environments surrounding a film is equal
to unity.

However in overwhelming majority of cases it is not so
\cite {Dressel}. As a rule in practice one deals with the films located on
some dielectric substrate. Meet also
cases when the metal film is located between two
dielectric environments. Generalization of the available theory of
 electromagnetic radiation interaction with
metal film on a such situation
will be the purpose of our work.

Thus, we consider a situation, in which
halfspace from which the electromagnetic wave falls on
film, has dielectric permeability $\varepsilon_1$.
We will consider that halfspace, in which electromagnetic
wave gets, leaving a film, has the dielectric
permeability $\varepsilon_2$. Last halfspace
is called as a substrate.

\begin{center}
{\bf 2. Statement of problem}
\end{center}

Let us consider the thin layer of metal located between two dielectric
environments. We will assume, that these environments are not magnetic.
Their dielectric permeability we will designate through $\varepsilon_1$
and $\varepsilon_2$.
Let us designate these environments the first and the second media accordingly.
We will assume that the first media are not absorbing.
Let on a film from the first media the electromagnetic wave falls.
Incidence angle we will designate as $\theta$.
Let us assume, that a vector of magnetic field of the electromagnetic
waves is parallel to a layer surface. Such wave is called E --
wave \cite {K} (or P -- wave \cite {F69}).

We take the Cartesian system of coordinates with the beginning of
coordinates on the surface of a layer adjoining to the first media.
Axis $x$ we will direct into the metal layer.
Axis $y$ we will direct  parallel to magnetic field vector
of electromagnetic wave.

The components of electric and magnetic field vectors we will
search in the form
$$
H_y(x,z,t)=H_y(x)e^{-i\omega t+ik_zz},
$$
and
$$
E_x(x,z,t)=E_x(x)e^{-i\omega t+ik_zz},\qquad
E_z(x,z,t)=E_z(x)e^{-i\omega t+ik_zz}.
$$

We denote the thickness of the layer by $d$.

The values $Z^{(1)}$ and $Z^{(2)}$ are designations of
impedancies on the bottom
layer surfaces   to antisymmetric on electric field
configurations of external fields (case 1) and symmetric configurations (case 2) accordingly.

In the case 1 we have the following relations on components of
electric and magnetic field and on derivative of electric
field
$$
E_z(0)=-E_z(d),\qquad H_y(0)=H_y(d), \qquad
\frac{d E_z(+0)}{d x}=\frac{d E_z(d-0)}{d x}.
\eqno{(1)}
$$

Accordingly in case 2 we have the following relations
$$
E_z(0)=E_z(d),\qquad H_y(0)=-H_y(d), \qquad
\frac{d E_z(+0)}{d x}=-\frac{d E_z(d-0)}{d x}.
\eqno{(2)}
$$

Let's notice, that last two relations on derivatives of
electric field from the first (1) and second (2) cases are
consequences of uniformity of the film.

Out of the layer it is possible to present electric fields in
the following form
$$
E_z^{(j)}(x)=\left\{\begin{array}{ll}
a_{j}h_{j}e^{ik_{2x}(x-d)}+b_{j}h_{j}e^{-ik_{2x}(x-d)},&
x>d, \\
h_{j}e^{ik_{1x}x}+p_{j}h_{j}e^{-ik_{1x}x},&
x<0, \quad j=1,2.
\end{array}
\right.
\eqno{(3)}
$$

Indexes "1" \, and "2" \, at factors $a_j, b_j, h_j, p_j$ and
field projections $E_y$ and $H_y$ correspond to the first
and to the second cases accordingly.

The impedance thus in both cases is defined as follows
(see, for example, \cite{S} and \cite{Landau8})
$$
Z^{(j)}=\dfrac{E_z^{(j)}(+0)}{H_y^{(j)}(+0)},\qquad j=1,2.
\eqno{(4)}
$$

\begin{center}
  \bf 3. Surface impedance (wave interaction with film)
\end{center}

From Maxwell equations we have following relations
for impedancies \cite{K}
$$
\frac{d E_z}{d x}-i\dfrac{ \omega}{c}\sin \theta E_x+
i\dfrac{ \omega}{c}H_y=0.
\eqno{(5)}
$$

Here $c$ is the speed of light.

For dielectric environments from Maxwell equations follows the
following relations \cite{K}
$$
i\dfrac{ \omega}{c} \varepsilon_{j} E_x-
i\dfrac{ \omega}{c}\sin \theta H_y=0, \qquad j=1,2.
\eqno{(6)}
$$

From last equations (6) we obtain on boundary of plasma
$$
E_x(-0)=\dfrac{\sin \theta}{\varepsilon_{1}}H_y(0),\qquad
E_x(d+0)=\dfrac{\sin \theta}{\varepsilon_{2}}H_y(d).
\eqno{(7)}
$$

Indexes "1" \, and "2" \, at dielectric permeability correspond
to the cases of two dielectric environments.

From the equation (5) and relations (7) we have
$$
\frac{d E_z}{d x}(-0)= -i\dfrac{ \omega}{c}\beta_1H_y(0),\qquad
\frac{d E_z}{d x}(d+0)= -i\dfrac{ \omega}{c}\beta_2H_y(d).
\eqno{(8)}
$$

Here
$$
\beta_{j}=1-\dfrac{\sin^2 \theta}{\varepsilon_{j}},\qquad j=1,2.
$$

Considering a continuity on border of plasma of quantities $E_z$ and
$H_y$ thus for quantity of an impedance (4) we receive \cite {Landau8}
$$
Z^{(j)}=
-i\dfrac{\beta_1\omega}{c}\dfrac{E_z^{(j)}(-0)}
{\dfrac{dE_z^{(j)}(-0)}{dx}}, \qquad j=1,2.
$$

The account of symmetry of the electric field for the first case according
to (1) and (3) leads to the following relations
$$
-a_1-b_1=1+p_1,
$$
$$
\beta_1k_{2x}(a_1-b_1)=
\beta_2k_{1x}(1-p_1).
$$

Solving this system, we get
$$
a_1=\dfrac{\beta_{2}k_{1x}-\beta_{1}k_{2x}}{2\beta_{1}k_{2x}}-
\dfrac{\beta_{2}k_{1x}+\beta_{1}k_{2x}}{2\beta_{1}k_{2x}}p_1,
$$
$$
b_1=-\dfrac{\beta_{2}k_{1x}+\beta_{1}k_{2x}}{2\beta_{1}k_{2x}}+
\dfrac{\beta_{2}k_{1x}-\beta_{1}k_{2x}}{2\beta_{1}k_{2x}}p_1.
$$

The account of symmetry of a field for the second case according
to (2) and (3) leads to the relations
$$
a_2+b_2=1+p_2,
$$
$$
\beta_1k_{2x}(a_2-b_2)=
-\beta_2k_{1x}(1-p_2).
$$

The solution of last system has the following form
$$
a_2=-\dfrac{\beta_{2}k_{1x}-\beta_{1}k_{2x}}{2\beta_{1}k_{2x}}+
\dfrac{\beta_{2}k_{1x}+\beta_{1}k_{2x}}{2\beta_{1}k_{2x}}p_2,
$$
$$
b_2=\dfrac{\beta_{2}k_{1x}+\beta_{1}k_{2x}}{2\beta_{1}k_{2x}}-
\dfrac{\beta_{2}k_{1x}-\beta_{1}k_{2x}}{2\beta_{1}k_{2x}}p_2.
$$

Let us consider the following configuration of the field
$$
E_y(x)=b_2h_2E_y^{(1)}(x)-b_1h_1E_y^{(2)}(x).
$$

Then the field $E_y(x)$ has the following structure
$$
E_y(x)=\left\{\begin{array}{ll}
(a_1b_2-a_2b_1)h_1h_2e^{ik_{2x}(x-d)},&
x>d, \\
(b_2-b_1)h_{1}h_2e^{ik_{1x}x}+(p_{1}b_2-p_2b_1)h_{1}h_2e^{-ik_{1x}x},&
x<0.
\end{array}
\right.
$$

Thus, the field corresponds to the electromagnetic wave falling on
the film from negative halfspace. A wave partially
passes through the film, and it is partially reflected.

Thus
$$
a_1b_2-a_2b_1=\dfrac{\beta_{2}k_{1x}}{\beta_{1}
k_{2x}}(p_2-p_1),
$$
$$
b_2-b_1=\dfrac{\beta_{2}k_{1x}+\beta_{1}k_{2x}}{\beta_{1}k_{2x}}+
\dfrac{\beta_{1}k_{2x}-\beta_{2}k_{1x}}{2\beta_{1}k_{2x}}(p_1+p_2),
$$
$$
p_1b_2-p_2b_1=
\dfrac{(p_1+p_2)(\beta_{2}k_{1x}+\beta_{1}k_{2x})}{2\beta_{1}k_{2x}}+
\dfrac{\beta_{1}k_{2x}-\beta_{2}k_{1x}}{\beta_{1}k_{2x}}p_1p_2.
$$

\begin{center}
  \bf  4. Transmittance, reflectance and absorptance
\end{center}

The quantities $p_1$ and $p_2$ may be to presented as
$$
p_{j}=\dfrac{ck_{1x}Z^{(j)}+ \beta_{1}\omega}{ck_{1x}Z^{(j)}-
\beta_{1}\omega},\qquad j=1,2.
$$

For reflection coefficient we obtain the following expression
$$
R=\Bigg|\dfrac{(p_1+p_2)(\beta_{2}k_{1x}+
\beta_{1}k_{2x})+2(\beta_{1}k_{2x}-\beta_{2}k_{1x})p_1p_2}
{2(\beta_{2}k_{1x}+\beta_{1}k_{2x})+
(\beta_{1}k_{2x}-\beta_{2}k_{1x})(p_1+p_2)}\Bigg|^2.
$$

The quantity $k_{2x}$ we can express as follows
\cite {Landau8}
$$
k_{2x}=\dfrac{ \omega}{c}
\sqrt{\varepsilon_2-\varepsilon_1\sin^2 \theta},
\qquad k_{1x}=\dfrac{ \omega}{c}\sqrt{\varepsilon_1}\cos \theta.
$$

Taking into account expressions for  $p_{1} $ and $p_2$
we can present them in the forms
$$
p_{j}=\dfrac{\sqrt{\varepsilon_1}\cos \theta
Z^{(j)}+ \beta_{1}}{\sqrt{\varepsilon_1}\cos \theta Z^{(j)}-
\beta_{1}},\qquad j=1,2.
$$

Time average  value of a flux of the energy of electromagnetic
fields $\langle{\bf S}\rangle$ is equal to \cite{Kizel}
$$
\langle{\bf S}\rangle=\dfrac{c}{16\pi}
\big\{[{\bf E}{\bf H}^*]+[{\bf E}^*{\bf H}]\big\}.
$$

Here the asterisk designates complex conjugation.

Using relation (8) for quantity $\langle{S_x}\rangle$ we get
$$
\langle{ S_x}\rangle_{j}=-\dfrac{c}{16\pi }(E_zH_y^*+E_z^*H_y)=
\dfrac{c^2}{8\pi \omega}|E_y|^2\Re\big(\dfrac{k_{jx}}{\beta_{j}}\big),
\qquad j=1,2.
$$

Let us enter a designation
$$
\bar{p}=\dfrac{p_1+p_2}{2}.
$$

We obtain for reflection coefficient the following result
$$
R=\Bigg|\dfrac{\beta_{12}\sqrt{\varepsilon_2-\varepsilon_1\sin^2 \theta}
(\bar p+p_1p_2)+\sqrt{\varepsilon_1}\cos \theta
(\bar p-p_1p_2)}{\beta_{12}\sqrt{\varepsilon_2-
\varepsilon_1\sin^2 \theta}
(1+\bar p)+\sqrt{\varepsilon_1}\cos \theta
(1-\bar p\Big)}\Bigg|^2.
\eqno{(9)}
$$

Here
$$
\beta_{12}=\dfrac{\beta_1}{\beta_2}=\dfrac{\varepsilon_2(\varepsilon_1-
\sin^2 \theta)}{\varepsilon_1(\varepsilon_2-\sin^2\theta)}.
$$

We can present the coefficient $T$ in the form
$$
T=\Re\Big(\beta_{12}\dfrac{k_{2x}}
{k_{1x}}\Big)\Big|\dfrac{a_1b_2-a_2b_1}{b_2-b_1}\Big|^2.
$$

We get now with using the obtaining relations
$$
T=4k_{1x}|
\Re\big(k_{2x}\beta_{12}\big)\Big|\dfrac{p_2-p_1}
{2(k_{1x}+\beta_{12}k_{2x})+
(\beta_{12}k_{2x}-k_{1x})(p_1+p_2)}\Big|^2,
$$
or
$$
T=\Re\Big\{\beta_{12}\sqrt{\varepsilon_1(\varepsilon_2-
\varepsilon_1\sin^2 \theta)}\Big\}\,\cos \theta
$$
$$
\times\left|\dfrac{p_2-p_1}
{\beta_{12}\sqrt{\varepsilon_2-
\varepsilon_1\sin^2 \theta}
(1+\bar p)+\sqrt{\varepsilon_1}\cos \theta
(1-\bar p\Big)}\right|^2.
$$

For transparent environments the quantities $\varepsilon_1$ and
$\varepsilon_2$ are real. From the obtained formula for the
transmission coefficient is clear, that in this case at
$$
\sin^2\theta \ge\dfrac{\varepsilon_2}{\varepsilon_1}
$$
the transmission coefficient is equal to zero, as
$$
\Re\Big\{\beta_{12}\sqrt{\varepsilon_1(\varepsilon_2-
\varepsilon_1\sin^2 \theta)}\,\Big\}=0.
$$

It corresponds to full internal reflection.

We note that by $\sin^2 \theta\to\varepsilon_2/\varepsilon_1$
the transmission coefficient  $T\to 0$.
The reflection coefficient  $R$ by $\theta\to \pi/2$ tends to $1$.

Now we can find the absorption coefficient  $A$ according to the formula
$$
A=1-T-R.
\eqno{(10)}
$$

Coefficients of transmission $T$ and reflection $R$ of the electromagnetic
waves by layer at $\varepsilon_1\to 1, \varepsilon_2\to 1$
transform in earlier known expressions \cite{F69}, \cite{F66}
$$
T=\dfrac{1}{4}\big|p_{1}-p_{2}\big|^2,\qquad
R=\dfrac{1}{4}\big|p_{1}+p_{2}\big|^2.
$$

Let us consider a case when electrons are specular reflected from a film
surface. Then for quantities $Z^{(j)}\;(j=1,2)$ are satisfied
the following relations \cite{F69}, \cite{F66}
$$
Z^{(j)}=-\dfrac{2i\Omega}{W}
\sum\limits_{n=-\infty}^{n=\infty}\dfrac{1}{Q^2}
\Big(\dfrac{Q_z^2}{\Omega^2\varepsilon_{l}}+
\dfrac{Q_x^2}{\Omega^2\varepsilon_{tr}-Q^2}\Big),\qquad j=1,2,
$$
where
$$
W=W(d)=\dfrac{\omega_pd}{c}\cdot 10^{-7},
$$
and the thickness of a film $d$ is measured in nanometers,
for $Z^{(1)}$ summation is made on odd $n$, and for
$Z^{(2)}$ on the even.

Here $\varepsilon_{tr}$ and $\varepsilon_l$ are  transverse and
longitudinal dielectric permeability accordingly, $\omega_p$ is the plasma
(Langmuir) frequency,
$$\varepsilon_{tr}=\varepsilon_{tr}(q_1, \Omega),\qquad
\varepsilon_l=\varepsilon_l(q_1,\Omega), \qquad \Omega=
\dfrac{\omega}{\omega_p},
$$
$$
\mathbf{q}_1=\dfrac{v_F}{c}\mathbf{Q}, \quad
\mathbf{Q}=\Big(Q_x,0,Q_z\Big), \quad
Q_x=\dfrac{\pi n}{W(d)}, \quad
Q_z=\sqrt{\varepsilon_1}\Omega\sin \theta,
$$
module of vector  $\mathbf{q}_1$ is equal to
$$
q_1=\dfrac{v_F}{c}\sqrt{\dfrac{\pi^2n^2}{W^2(d)}+\varepsilon_1\Omega^2
\sin^2 \theta},
$$
$$
\varepsilon_{tr}(q_1,\Omega)=1-\dfrac{3}{4\Omega q_1^3}\Bigg[
2(\Omega+i\varepsilon)q_1+\Big[(\Omega+i\varepsilon)^2-
q_1^2\Big]\ln\dfrac{\Omega+i\varepsilon-q_1}
{\Omega+i\varepsilon+q_1}\Bigg],
$$
$$
\varepsilon_l(q_1, \Omega)=1+\dfrac{3}{q_1^2}
\dfrac{1+\dfrac{\Omega+i\varepsilon}{2q_1}
\ln\dfrac{\Omega+i\varepsilon-q_1}{\Omega+i\varepsilon+
q_1}}{1+\dfrac{i\varepsilon}{2q_1}
\ln\dfrac{\Omega+i\varepsilon-q_1}{\Omega+i\varepsilon+q_1}},
$$
$\mathbf{q}_1$ is the dimensionless wave vector, $\mathbf{q}=
\dfrac{\omega_p}{v_F}\mathbf{q}_1$ is the dimensional wave vector,
$\varepsilon=\dfrac{\nu}{\omega_p}$, $\nu$ is the effective
electron collision frequency.

We transform now these functions $Z^{(1)}$ and $Z^{(2)}$
$$
Z^{(1)}=-\dfrac{4i\Omega}{W(d)}\sum\limits_{n=1}^{+\infty}
\dfrac{1}{\Omega^2
\varepsilon_{tr}(\Omega,\varepsilon,d,2n-1,\theta,\varepsilon_1)-
Q(\Omega,d,2n-1,\theta,\varepsilon_1)},
$$
$$
Z^{(2)}=-\dfrac{2i\Omega}
{W(d)\Big[\Omega^2
\varepsilon_{tr}(\Omega,\varepsilon,d,0,\theta,\varepsilon_1)-
Q(\Omega,d,0,\theta,\varepsilon_1)}-
$$
$$
-\dfrac{4i\Omega}{W(d)}\sum\limits_{n=1}^{+\infty}
\dfrac{1}{\Omega^2
\varepsilon_{tr}(\Omega,\varepsilon,d,2n,\theta,\varepsilon_1)-
Q(\Omega,d,2n,\theta,\varepsilon_1)}.
$$

We will consider the important special case of formula
for transmission coefficient. We will assume that
$\varepsilon_1$ and $\varepsilon_2$ are real.
Then expression for transmission coefficient
we can present in the following form
$$
T=\cos\theta \beta_{12}
\Re\sqrt{\varepsilon_1(\varepsilon_2-\varepsilon_1
\sin^2\theta)}
\Bigg|\dfrac{p_1-p_2}{\beta_{12}\sqrt{\varepsilon_2-
\varepsilon_1\sin^2\theta}(1+\bar p)+\sqrt{\varepsilon_1}\cos\theta
(1-\bar p)}\Bigg|^2.
\eqno{(11)}
$$

We will introduce the new quantity
$
\varepsilon_{12}=\dfrac{\varepsilon_2}{\varepsilon_1}.
$
Now the formuas (11) and (9) transform to the form
$$
T=\beta_{12} \cos\theta \Re\sqrt{\varepsilon_{12}-\sin^2\theta}
\Bigg|\dfrac{p_1-p_2}{\beta_{12}\sqrt{\varepsilon_{12}-
\sin^2\theta}(1+\bar p)+\cos\theta
(1-\bar p)}\Bigg|^2,
\eqno{(12)}
$$
and
$$
R=\Bigg|\dfrac{\beta_{12}\sqrt{\varepsilon_{12}-\sin^2\theta}
(\bar p+p_1p_2)+\cos\theta(\bar p-p_1p_2)}
{\beta_{12}\sqrt{\varepsilon_{12}-\sin^2\theta}(1+\bar p)+\cos\theta
(1-\bar p)}\Bigg|^2.
\eqno{(13)}
$$

\begin{center}
  \bf 5. Analysis of results
\end{center}

Let us consider the case of a thin film of sodium.
Then \cite{F69}
$\omega_p=6.5\times 10^{15}\sec^{-1}$, $v_F=8.52\times 10^7$
$\rm cm/\sec$.

We will carry out graphic analysis of coefficients of
transmission, reflections and absorption.
We will use formulas (12) and (13) for coefficients of
transmission and reflections, and coefficient of absorption we
will analyze using the formula (10).

On fig. 1--6 we represent the behavior of
transmission coefficient (Fig. 1--3) and reflection coefficient
(Fig. 4--6) as functions of dimensionless frequency
of an electromagnetic wave $\Omega$.
The incidence angle of an electromagnetic wave is equal to
$\theta=75^\circ$,
dimensionless frequency of electron collisions is equal to
$\varepsilon=10^{-3}$, i.e. dimensional frequency of collisions is
equal to $\nu=0.001\omega_p $.

Comparison with similar result \cite{LY2011}
for interaction of electromagnetic p--waves with freely hanging
thin film shows, that substrate influence leads
to some reduction of transmittance and increase of reflectance.
In the region of resonant frequencies the behavior of
coefficients of transmission and reflection somewhat varies.

We consider now the system glass -- metal film -- air.
We take then $\varepsilon_1=4, \varepsilon_2=1$.
On Fig. 7 we  represent behavior of all coefficients of
transmission, reflection and absorption as functions of a
incidence angle of the electromagnetic wave.
For the film  thickness  10 nanometers the transmittance
is the monotonously decreasing function, the reflectance is the
monotonously increasing function, and the absorptance has one
maximum.

On Fig. 8 we  represent the behavior of coefficients $T, R$ and $A$
as functions of dimensionless frequency
$\Omega \, (0\leqslant \Omega \leqslant 1.5)$.
We consider the system mica -- film -- air
($\varepsilon_1=8, \varepsilon_2=1$). The film thickness is equal to
100 nanometers. We take $\nu=0.001 \omega_p$, $\theta=15^\circ$.
It is interesting to notice, that the coefficients of transmission
and absorption have  minimum in the point $\Omega=1$, i.e.
at $\omega =\omega_p$, and
the reflection coefficient has the sharp maximum in this point.

On Fig. 9 we represent character of "comb teeth" of
transmission coefficient for the film  10 nanometers thickness
in region of the resonant frequencies
($1.4 \leqslant \Omega \leqslant 1.53$) for collision electron
frequency $\nu=0.001\omega_p$ and for a incidence angle
of electromagnetic wave $\theta=15^\circ $.
Thus the system mica -- film -- air, i.e.
$\varepsilon_1=8, \varepsilon_2=1$ is considered.

\begin{center}
  \bf 6. Conclusion
\end{center}

In the present work generalization of the theory of interaction
electromagnetic radiation with a metal film on the case of
a film concluded between two various dielectric environments is produced.
Formulas for coefficients of transmission, reflection and absorption
of an electromagnetic wave are deduced.
The graphic analysis of these coefficients depending on
oscillation frequency of an electromagnetic field,
incidence angle of an electromagnetic waves,
thickness of the film and character  of the second
dielectric environment (substrate).

\begin{figure}[h]\center
\includegraphics[width=16.0cm, height=6.5cm]{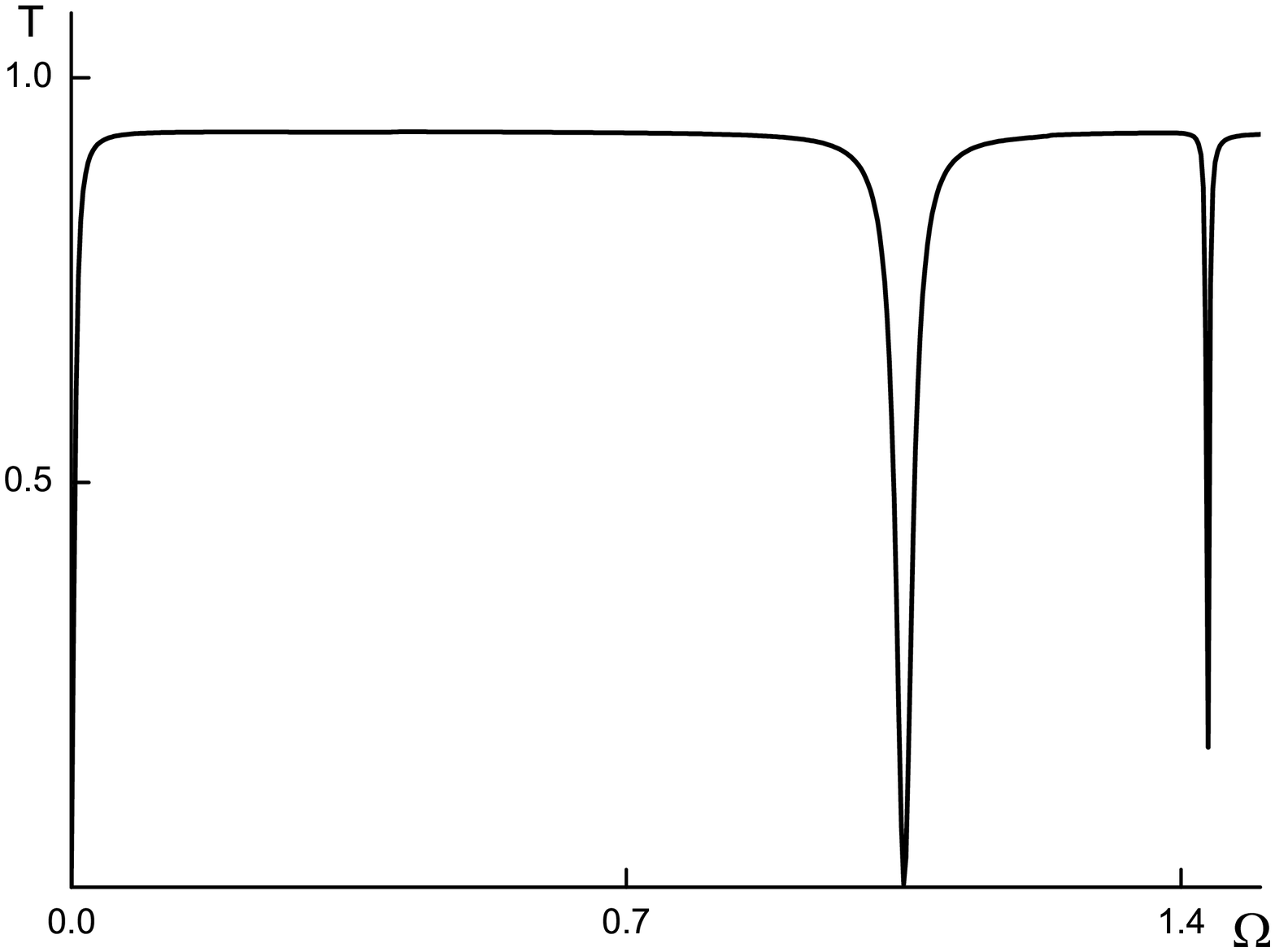}
\noindent\caption{Transmittance, air--film--glass, $d=1$  nm,
$0\leqslant \Omega \leqslant 1.5$,
$\nu=0.001\omega_p$, $\theta=75^\circ$.}
\includegraphics[width=16.0cm,  height=6.5cm]{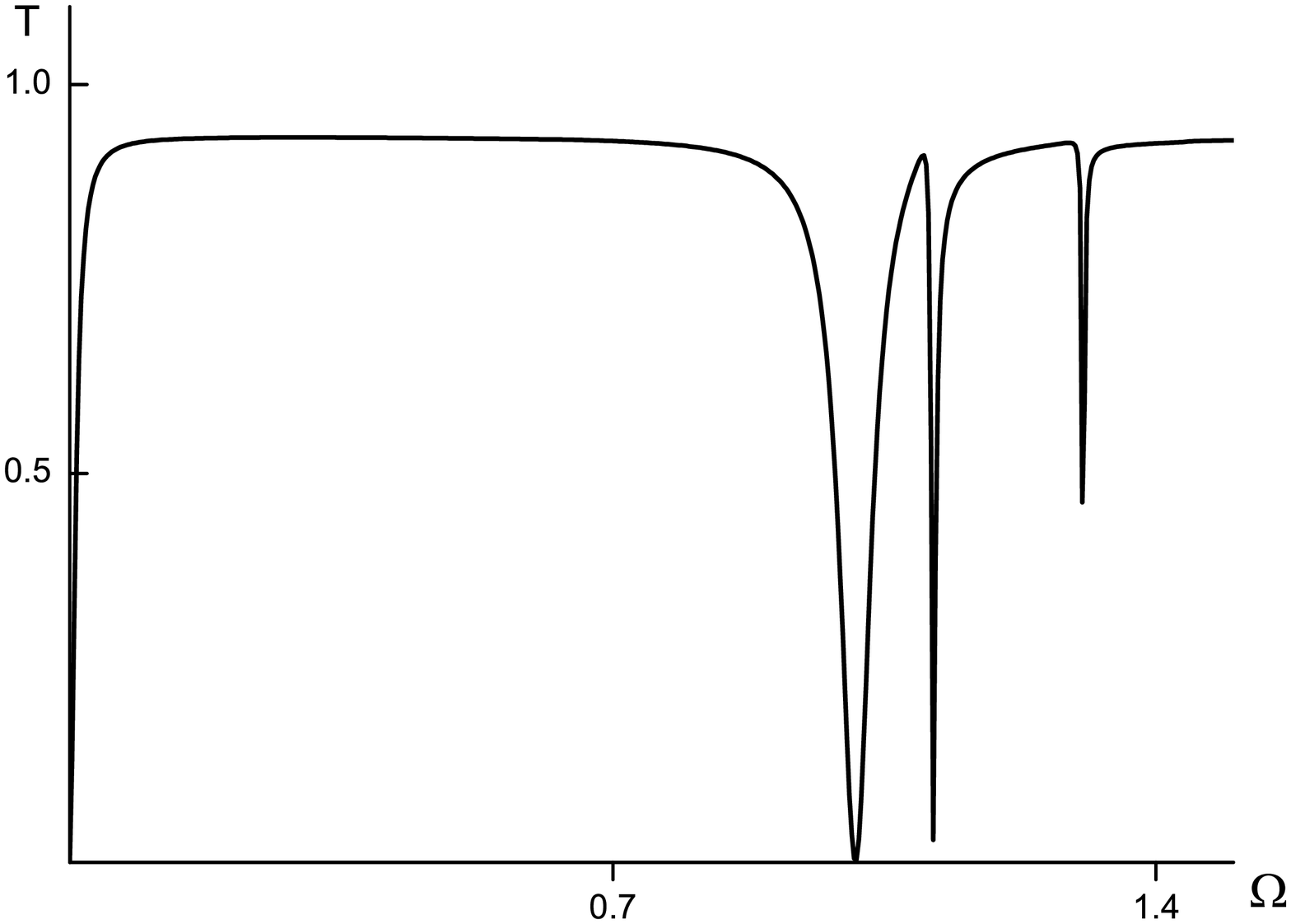}
\noindent\caption{Transmittance, air--film--glass, $d=2$  nm,
$0\leqslant \Omega \leqslant 1.5$,
$\nu=0.001\omega_p$, $\theta=75^\circ$.}
\includegraphics[width=16.0cm, height=6.5cm]{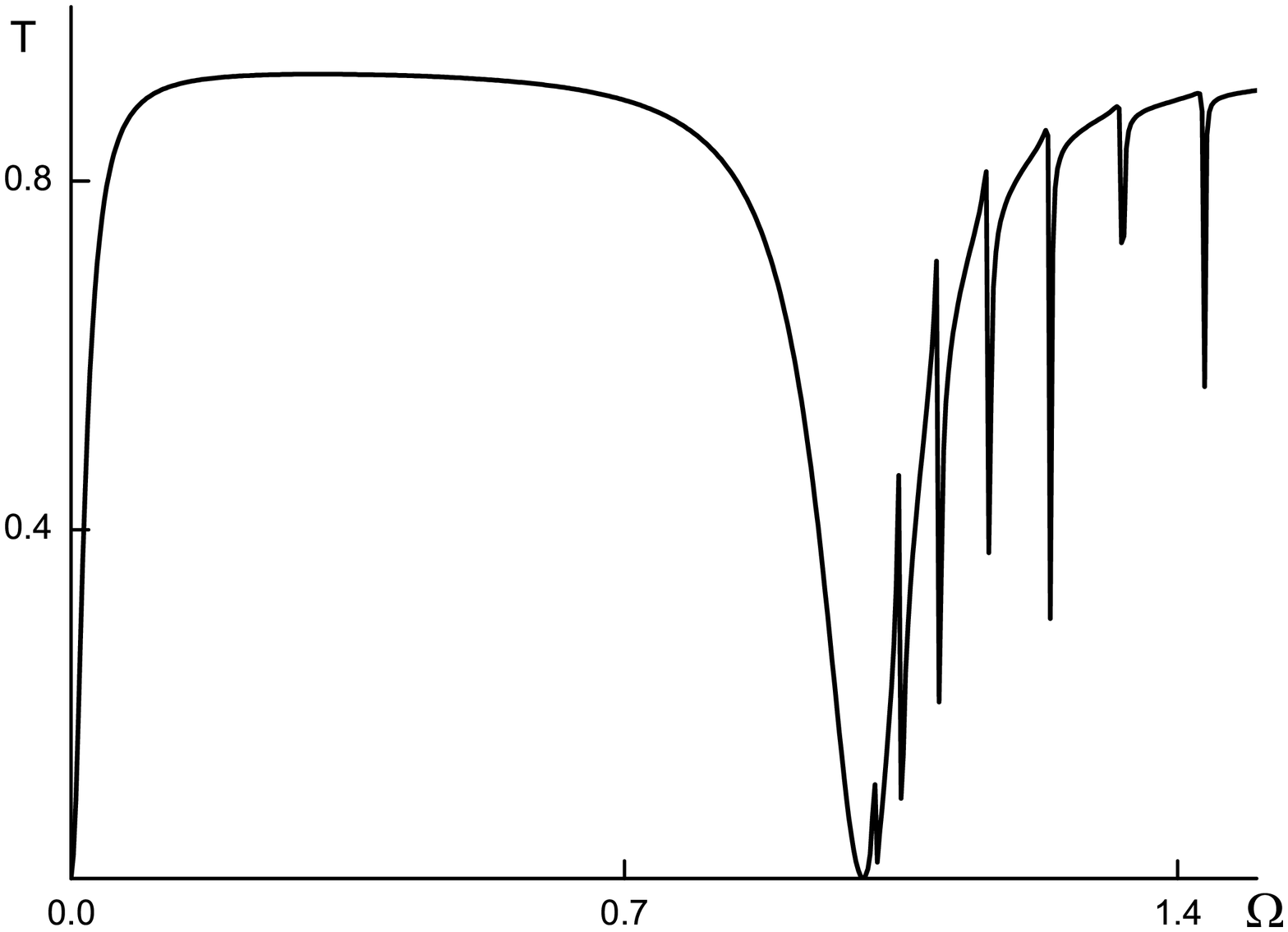}
\noindent\caption{Transmittance, air--film--glass, $d=5$  nm,
$0\leqslant \Omega \leqslant 1.5$,
$\nu=0.001\omega_p$, $\theta=75^\circ$.}
\end{figure}

\begin{figure}[t]\center
\includegraphics[width=16.0cm, height=6.5cm]{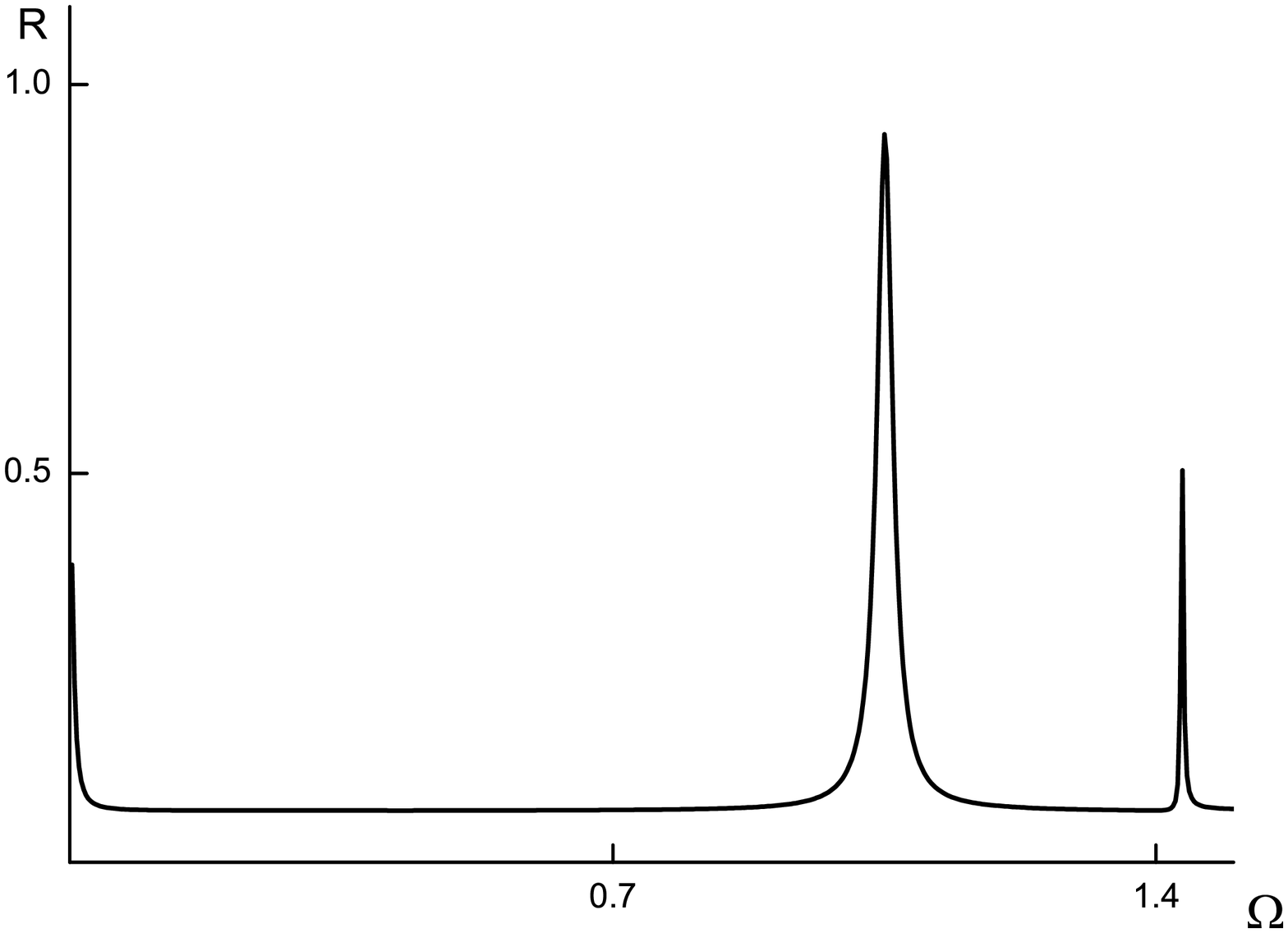}
\noindent\caption{Reflectance, air--film--glass, $d=1$  nm,
$0\leqslant \Omega \leqslant 1.5$,
$\nu=0.001\omega_p$, $\theta=75^\circ$.}
\includegraphics[width=16.0cm,  height=6.5cm]{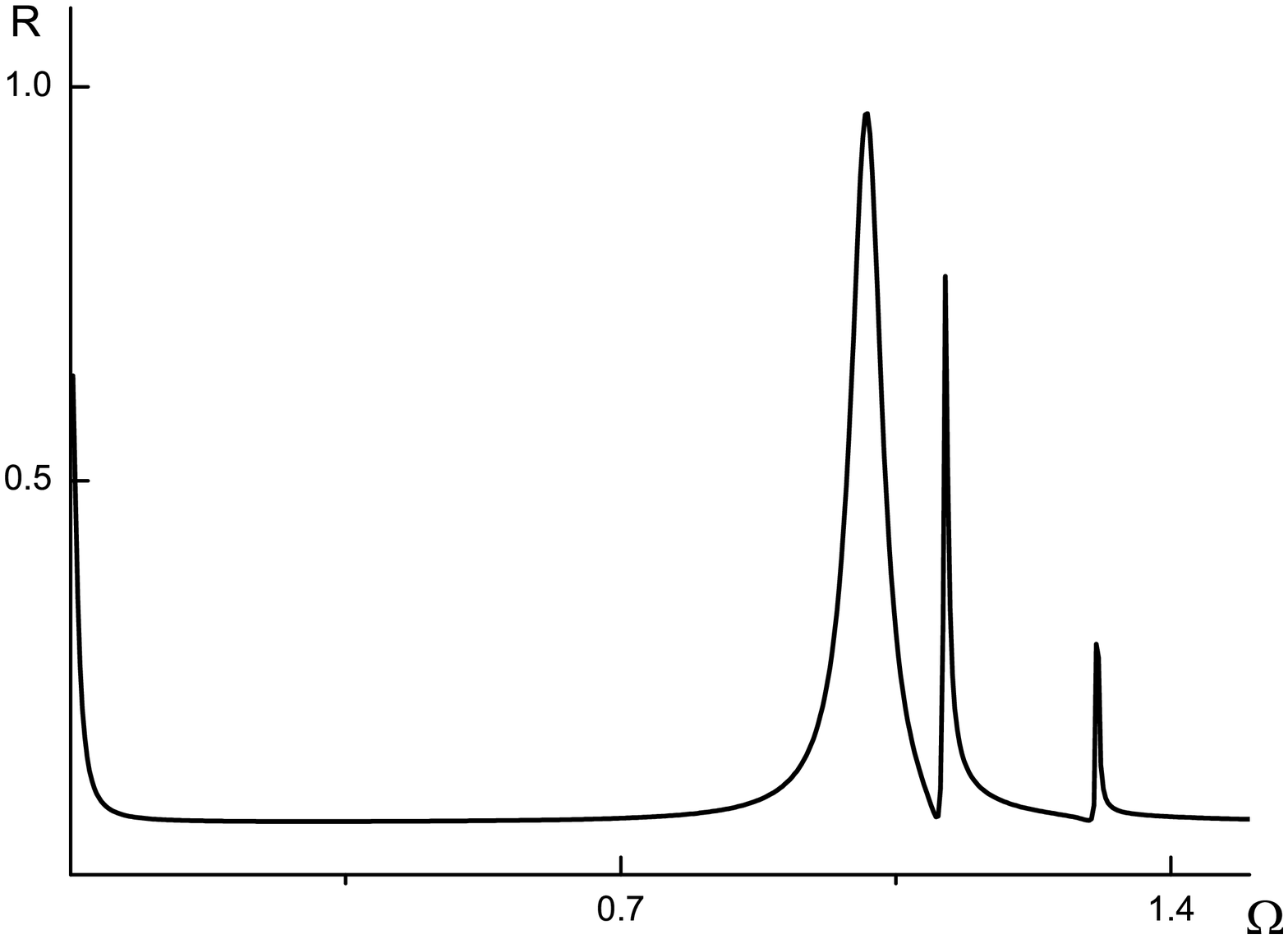}
\noindent\caption{Reflectance, air--film--glass, $d=2$  nm,
$0\leqslant \Omega \leqslant 1.5$,
$\nu=0.001\omega_p$, $\theta=75^\circ$.}
\includegraphics[width=16.0cm, height=6.5cm]{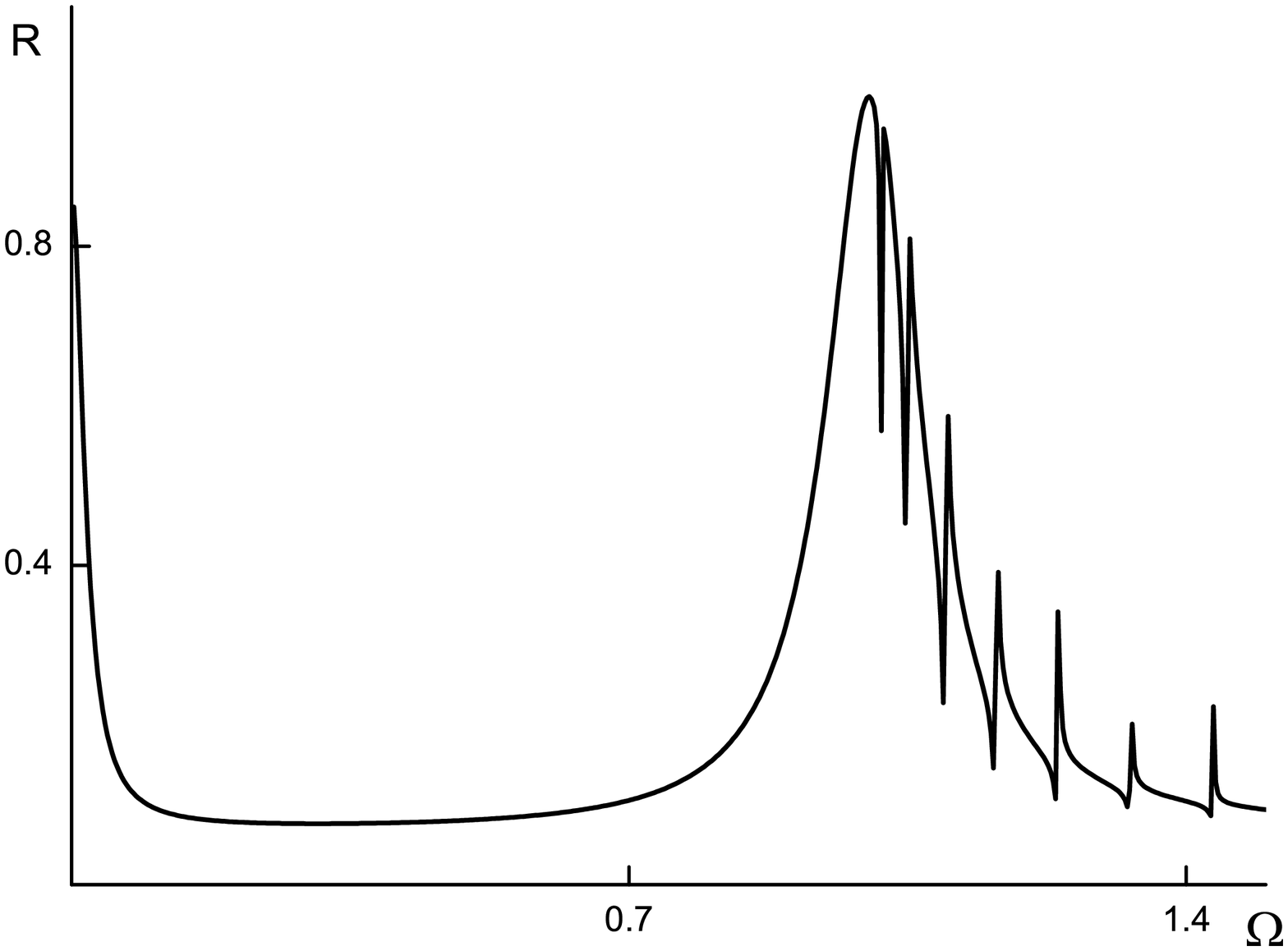}
\noindent\caption{Reflectance, air--film--glass, $d=5$  nm,
$0\leqslant \Omega \leqslant 1.5$,
$\nu=0.001\omega_p$, $\theta=75^\circ$.}
\end{figure}

\begin{figure}[t]\center
\includegraphics[width=16.0cm, height=6.5cm]{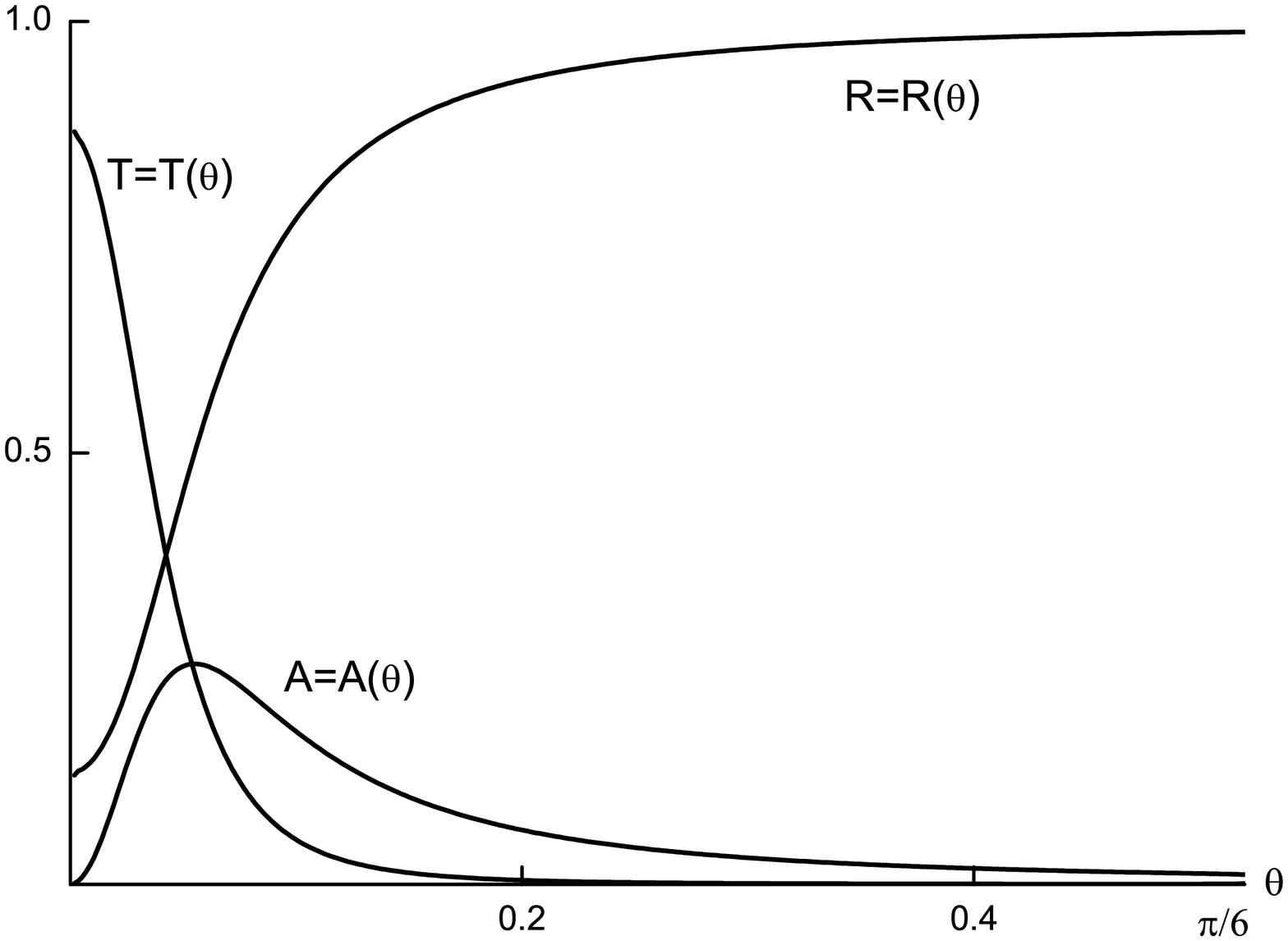}
\noindent\caption{Transmittance, Reflectance, Absorptance,
glass--film--air, $d=10$  nm,
$\Omega=1$, $\varepsilon_1=4,
\varepsilon_2=1$,
$\nu=0.001\omega_p$, $\theta=15^\circ$.}
\includegraphics[width=16.0cm,  height=6.5cm]{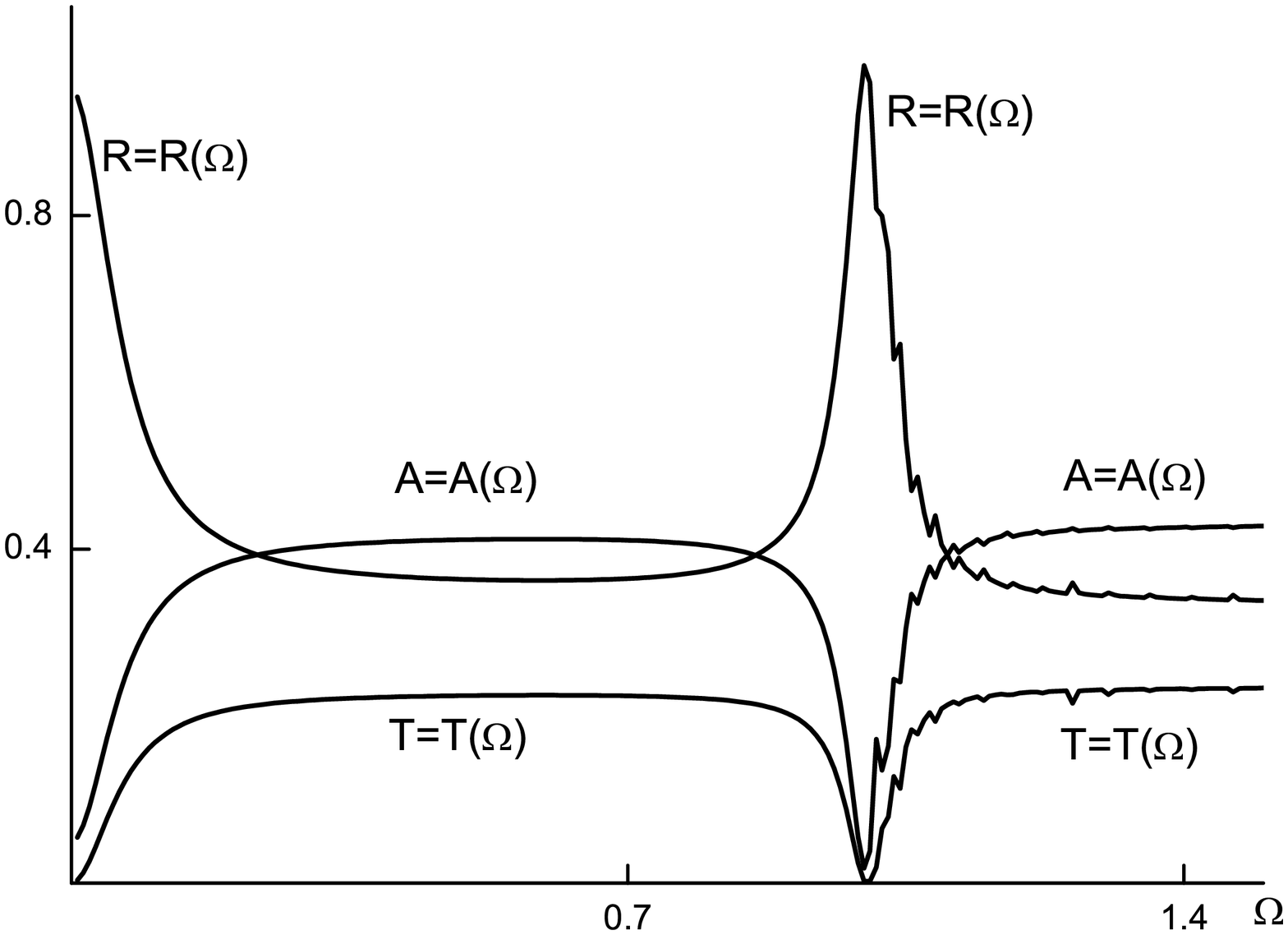}
\noindent\caption{Transmittance, Reflectance, Absorptance,
mica--film--air, $d=100$  nm,  $\varepsilon_1=8, \varepsilon_2=1$,
$0\leqslant \Omega \leqslant 1.5$,
$\nu=0.001\omega_p$, $\theta=15^\circ$.}
\includegraphics[width=16.0cm,  height=6.5cm]{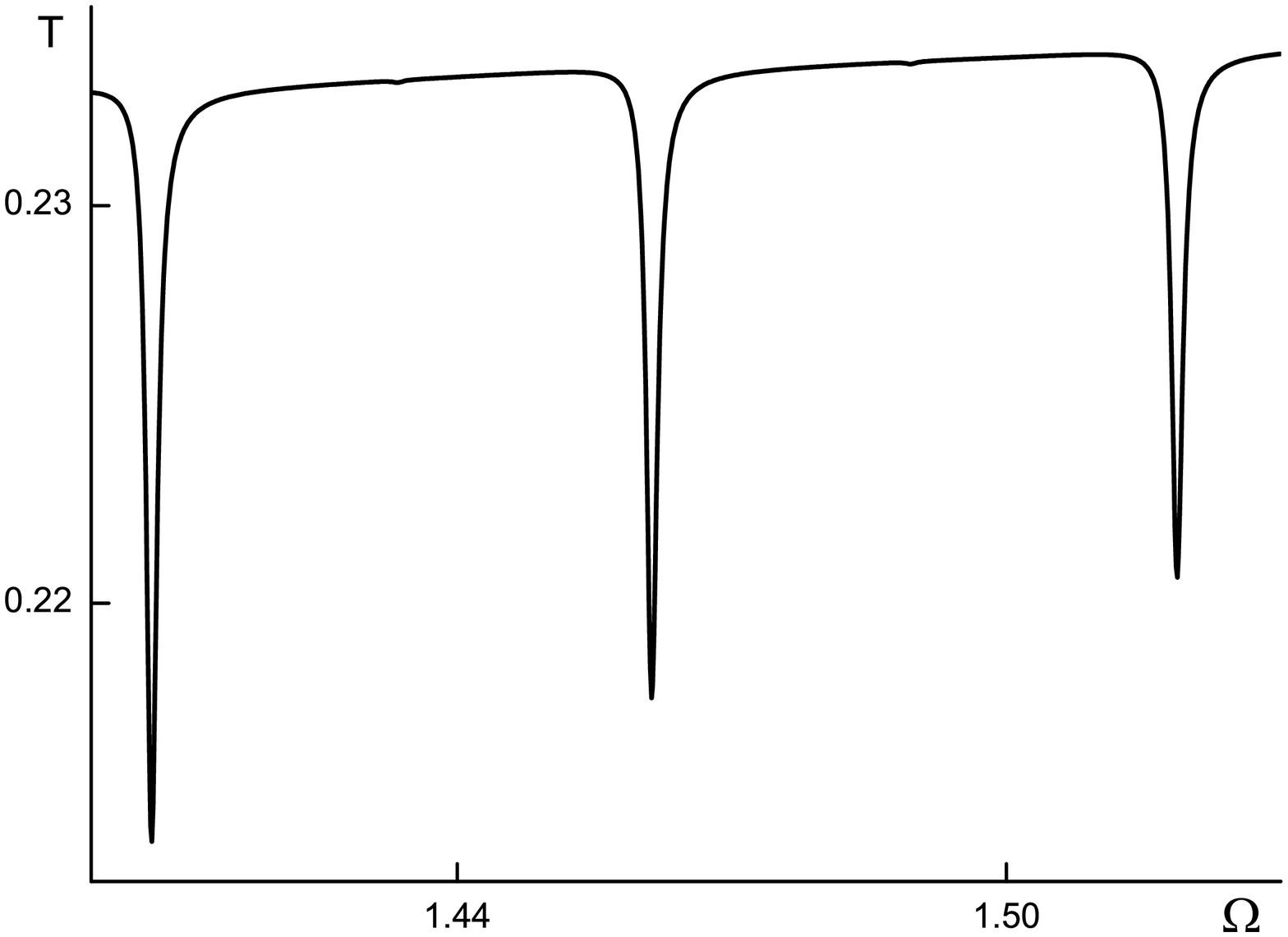}
\noindent\caption{Character of "comb teeth",
glass--film--air, $d=10$  nm, $1.4\leqslant \Omega \leqslant 1.53$,
$\nu=0.001\omega_p$, $\theta=15^\circ$.}
\end{figure}

\end{document}